\documentclass[useAMS,usenatbib]{mn2e}
\usepackage{epsfig}
\usepackage{float}

\usepackage{natbib}

\usepackage{color}

\usepackage{aas_macros}

\def\eSF{\mbox{$\epsilon_{\rm SF}$}}

\def\rhalf{\mbox{$r_{\rm 1/2}$}}

\def\Msun{\mbox{$M_\odot$}}

\def\mst{\mbox{$M_{\star}$}}

\def\lsim{\mathrel{\rlap{\lower3.5pt\hbox{\hskip0.5pt$\sim$}}
    \raise0.5pt\hbox{$<$}}}                
\def\gsim{~\rlap{$>$}{\lower 1.0ex\hbox{$\sim$}}}

\def\gZ{\mbox{$\nabla_{\rm Z}$}}

\def\gage{\mbox{$\nabla_{\rm age}$}}

\def\ggi{\mbox{$\nabla_{\rm g-i}$}}

\def\gVIa{\mbox{$\nabla_{\rm F606W-F814W}$}}
\def\gVIb{\mbox{$\nabla_{\rm F606W-F850LP}$}}

\def\gIIRb{\mbox{$\nabla_{\rm F850LP-F160W}$}}

\def\gVIc{\mbox{$\nabla_{\rm F555W-F814W}$}}
\def\gVId{\mbox{$\nabla_{\rm F450W-F814W}$}}

\def\gi{\mbox{$g-i$}}

\def\Eone{{\tt E1}}
\def\Etwo{{\tt E2}}
\def\Ethree{{\tt E3}}
\def\Efour{{\tt E4}}

\title[Colour gradients]{Colour gradients of high-redshift Early-Type Galaxies from hydrodynamical monolithic models}

\author[Tortora C. et al.]{\noindent
C.~Tortora$^{1}$\thanks{E-mail: ctortora@physik.uzh.ch},
A.~Pipino$^{2}$, A.~D'Ercole$^{3}$, N.R.~Napolitano$^{4}$,
F.~Matteucci$^{5,6}$
\\~\\
$^1$ Universit$\ddot{a}$t Z$\ddot{u}$rich, Institut f$\ddot{u}$r
Theoretische Physik, Winterthurerstrasse
190, CH-8057, Z$\ddot{u}$rich, Switzerland \\
$^2$ Institut f$\ddot{u}$r Astronomie, ETH Z$\ddot{u}$rich,
Wolfgang-Pauli-Str. 27, 8093
Z$\ddot{u}$rich, Switzerland\\
$^3$ INAF -- Osservatorio Astronomico di Bologna, via Ranzani 1,
40127 Bologna, Italy\\
$^4$ INAF -- Osservatorio Astronomico di Capodimonte, Salita
Moiariello, 16, 80131 - Napoli, Italy\\
$^5$ Dipartimento di Fisica, Sez. Astronomia, Universit\'a di
Trieste, via G.B.Tiepolo 11, 34100, Trieste, Italy\\
$^6$ INAF Osservatorio Astronomico di Trieste, via G.B.Tiepolo 11,
34100, Trieste, Italy\\}

\begin{document}
\date{Accepted  Received }
\pagerange{\pageref{firstpage}--\pageref{lastpage}} \pubyear{xxxx}
\maketitle

\label{firstpage}
\begin{abstract}

We analyze the evolution of colour gradients predicted by the
hydrodynamical models of early type galaxies (ETGs) in
\cite{Pipino+08}, which reproduce fairly well the chemical
abundance pattern and the metallicity gradients of local ETGs. We
convert the star formation (SF) and metal content into colours by
means of stellar population synthetic model and investigate the
role of different physical ingredients, as the initial gas
distribution and content, and \eSF, i.e. the normalization of SF
rate.  From the comparison with high redshift data, a full
agreement with optical rest-frame observations at $z \lsim 1$ is
found, for models with low \eSF, whereas some discrepancies emerge
at $1 < z < 2$, despite our models reproduce quite well the data
scatter at these redshifts. To reconcile the prediction of these
high \eSF\ systems  with the shallower colour gradients observed
at lower z we suggest intervention of 1-2 dry mergers. We suggest
that future studies should explore the impact of wet galaxy
mergings, interactions with environment, dust content and a
variation of the Initial Mass Function from the galactic centers
to the peripheries.
\end{abstract}


\section{Introduction}\label{sec:intro}

Radial variations in ages and metallicities of stellar populations
(SPs) have been shown to be efficient tools to discriminate among
different galaxy formation scenarios. Recently, a plethora of
studies have dedicated attention to the analysis of gradients in
colours and absorption features for wide samples of local
early-type galaxies (ETGs; e.g. \citealt{Spolaor+09};
\citealt{Kuntschner+10}; \citealt{Rawle+10}; \citealt{Spolaor+10};
\citealt[herafter T+10]{Tortora+10CG}; \citealt{Tortora+11CGsim}),
deriving SP gradients by means of stellar population synthesis
(SPS) analysis. Whilst these works consistently find a decrease in
the metallicity at large radii as the cause of both line-strength
and colour gradients, the scatter in the galaxy properties at
fixed stellar mass still prevent from deriving robust constraints
on the competing galaxy formation models on the market
(\citealt{Pipino+10}).

For instance, a well known class of these models is represented by
the so-called monolithic models. Steep metallicity gradients
($\sim -0.5 \, \rm dex$ per decade variation in radius) are
expected from classical dissipative collapse models (e.g.
\citealt{Larson74}; \citealt{Carlberg84}) and their revised
versions starting from semi-cosmological initial conditions (e.g.
\citealt{Kawata01}, \citealt{Kobayashi04}). These metallicity
gradients arise because the stars form everywhere in the
collapsing cloud and then remain in orbit with a little inward
motion whereas the gas falls down because of dissipation. This
falling gas contains the new metals ejected by evolving stars so
that a metallicity gradient develops in the gas. As stars continue
to form, their composition reflects the gaseous abundance
gradient. \cite{Pipino+08} and \cite{Pipino+10} have shown that
monolithic hydrodynamical models can be able to satisfy different
local correlations, as the mass--metallicity and the
mass--$[\alpha/Fe]$ relations and produce metallicity gradients of
$\sim -0.3 \, \rm dex$, consistent with several observational
samples (\citealt{Spolaor+09}; \citealt{Spolaor+10};
\citealt{Kuntschner+10}; \citealt{Rawle+10};
\citealt{Tortora+10CG}; \citealt{LaBarbera+11_CG}). These results
have been possible thanks to the requirement, fulfilled by the
models, that the stars produced have high average $[\alpha/Fe]$ in
the cores and thus short SF, and smaller age gradients.

It is however important to note that some simulations in the suite
by \cite{Pipino+08} and \cite{Pipino+10} still feature very steep
metallicity gradients of $\sim -0.5 \, \rm dex$, explaining the
observed scatter in massive ETGs (e.g. \citealt{Tortora+10CG}) in
terms of variations in the initial conditions and in particular in
the efficiency of star formation (SF) processes. These authors
explain the metallicity gradients in massive ETGs without
recurring to strong AGN feedback (\citealt{Tortora+09AGN};
\citealt{Tortora+10CG}), which have been suggested to flatten the
gradients. Cosmological simulations require that the formation of
massive galaxies must include at least a few ``dry mergers''
(\citealt{deLucia+06}): these would flatten any pre-existing
metallicity gradients (\citealt{Bekki_Shioya99};
\citealt{Kobayashi04}; \citealt{DiMatteo+09}). Only if the
pre-existing gradients in the progenitors of a merging system are
enough steep, it is possible that the descendant model galaxy will
match the observed values at $z\sim$0 (\citealt{CDB93};
\citealt{Davies+93}; \citealt{Sanchez-Blazquez+06};
\citealt{Sanchez-Blazquez+07}; \citealt{Annibali+07};
\citealt{Ogando+05}; \citealt{Spolaor+09};
\citealt{Tortora+10CG}). Clearly, observations of local massive
ellipticals do not have enough constraining power as models
improve in complexity, each of them featuring a mixture of
different channels for galaxy formation, In order to break such
degeneracies, both observational and numerical studies at
different environments and redshift are needed.

From the observational point of view, to have a full picture is
necessary to analyze colour and SP gradients in higher-redshift
galaxies. Unfortunately, the available datasets cover local
environments and only few studies allow the investigation of
colour and SP gradients at higher redshifts
(\citealt{LaBarbera+04}; \citealt{Tamura+00};
\citealt{Tamura_Ohta00}; \citealt{Menanteau+01};
\citealt{Ferreras+09_GOODS}; \citealt{Guo+11};
\citealt{Gargiulo+11,Gargiulo+12}; \citealt{Welikala_Kneib12}).

In the present paper we further exploit a sample of the
hydrodynamical models for massive ETGs in \cite{Pipino+08}, we
transform the SPs of the model predictions (evolution in metal
content and SF history) into magnitudes and colours using the SPS
in \cite{BC03}. This approach allows us to compare our models with
observed colour gradients and avoid typical systematics (as the
age-metallicity degeneracy) plaguing the SPS fitting. We will
concentrate on two different aspects:
\begin{enumerate}
\item the evolution with redshift of the colour gradients and the
analysis of the role of the initial conditions of the models, as
the initial gas density distribution, its mass content, and the
rate of SF processes;
\item the comparison with observed color gradients at low/moderate redshift ($z < 1$) and  high
redshift (i.e. $z > 1$),
\end{enumerate}
with the aim of testing the consequences of assuming one
particular model, and to make predictions to be validated with
future data.

The paper is organized as follows. In Sect. \ref{sec:setup} we
discuss the simulation setup and the main characteristics of the
models. The evolution with redshift of the colour gradients in the
rest-frame of the galaxies is shown in Sect.
\ref{sec:CG_evolution}, while the comparison with high redshift
observations and systematics are analyzed in Sect.
\ref{sec_observations}. In Sect. \ref{sec:discussion} we discuss
our results within a wider formational scheme and conclusions are
outlined in Sect. \ref{sec:conclusions}.

\begin{table*}
\centering \caption{Model input and output parameters, from
\citet{Pipino+08}. {\tt E}i codes identify model galaxies. Models
\Eone, \Etwo, \Ethree\ and \Efour\ correspond to {\tt Ma1}, {\tt
Mb3}, {\tt Mb4} and {\tt LB} in \citet{Pipino+08}, respectively.
Columns report, as imput parameters: 1) the name of the model, 2)
the initial central gas density $\rho_{\rm 0, gas}$, 3) the
initial gas profile, 4) the SF parameter \eSF\ (see
Eq.~(\ref{sfr})) and 5) the DM mass $M_{\rm DM}$, and the final
($z=0$) output parameters: 6) the stellar mass \mst, 7) the
half-mass radius \rhalf, 8) [O/Fe] of SPs in the galactic core, 9)
age gradient, \gage, 10) Z gradient, \gZ. Gradients in columns 9)
and 10) are calculated accordingly to \citet{Pipino+08} and
\citet{Pipino+10} as $\nabla_{X} = (\log X_{\rm core}- \log X_{\rm
e})/( \log r_{\rm core}- \log \rhalf)$, where $\rm X = \rm Z$,
$\rm age$, and $r_{\rm core}$ is set as in Table 2 in
\citet{Pipino+08}. The stellar and DM mass are limited to their
tidal radii (see \citet{Pipino+08} for details). $\flat$ Note
that, for \Efour, after fixing a bug, the local gradient getting
out is steeper than the one shown in \citet{Pipino+08,
Pipino+10}.}\label{tab:SIMs}
\begin{tabular}{lcccccccccc} \hline
& \multicolumn{4}{c}{Input} & & \multicolumn{4}{c}{Output} \\
\hline
Model & $\rho_{\rm 0, gas}$ & Initial profile & \eSF\  & $M_{\rm DM}$ & & \mst  & \rhalf & $[O/Fe]_{\rm core}$ & \gage &  \gZ  \\
{} & ($10^{-25} \, \rm g \, cm^{-3}$) & {} & {}  & ($10^{11}\, \rm \Msun$)& & ($10^{10} \Msun$) & (kpc) & (dex) & (dex) & (dex) \\
 \hline
\Eone   & 0.6  & IS   & 1  & 22 & & 6.0 & 12  & 0.29 & $2.8 \times 10^{-4}$ & -0.2 \\ 
\Etwo   & 0.06 & flat & 10 & 22 & & 21  & 8.8 & 0.17 & $2.3 \times 10^{-3}$ & -0.35 \\ 
\Ethree & 0.6  & flat & 1  & 22 & & 26  & 5.4 & 0.42 & $3.6 \times 10^{-4}$ & -0.21 \\ 
\Efour  & 0.6  & flat & 10 & 57 & & 29  & 21  & 0.12 & $6.0 \times 10^{-3}$ & -0.45$^{\flat}$ \\ 
\hline
\end{tabular}
\end{table*}

\section{Simulation setup and spectral synthesis}\label{sec:setup}

We adopted the hydrodynamical model in \cite{Pipino+08} and
\cite{Pipino+10} to predict the evolution of SF and metallicity
and SPS models from \cite{BC03} to link the predicted SP evolution
to observable quantities as colours. Below we briefly summarize
the simulation specifics and typical behaviours.

\subsection{Hydrodynamical model}

\subsubsection{General setup}

We use a one-dimensional hydrodynamical model that follows the
time evolution of the density of mass, momentum and internal
energy of a galaxy, under the assumption of spherical symmetry.
The standard gas-dynamical equations (\citealt{Bedogni_Dercole89},
\citealt{Pipino+08}) include source terms to describe the
injection of total mass and energy in the gas due to the mass
return and energy input from the stars, including SNIa and SNII in
a self-consistent way. This allows the code to follow the
hydrodynamical evolution of H, He, O and Fe in detail.

The grid in divided in 550 zones with a size ratio between
adjacent zones equal to 1.03. The innermost zone is 10 pc wide.
At the grid center we impose reflecting boundary condition,
whereas we set outflow condition in the
outermost point.

At every point of the mesh we allow the SF to occur at the
following rate:
\begin{equation}
\Psi = \nu  \rho = {\epsilon_{SF} \over max(t_{cool},t_{ff})} \rho
\, \label{sfr}
\end{equation}
where $t_{cool}$ and $t_{ff}$ are the \emph{local} cooling and
free-fall timescales, respectively, $\rho$ is the mass density and
$\epsilon_{SF}$ is a suitable {\it SF parameter} that contains all
the uncertainties on the timescales of the SF process. We assume
that the stars do not move from the gridpoints at which they have
been formed, since we expect that the stars will spend most of
their time close to their apocentre.

As previously discussed in \cite{Pipino+08}, very strong SN
feedback can stop SF processes too early, predicting high
$\alpha$-enhancement in the galactic core and a too diffuse
galaxy. On the other hand, the energy input in the ISM by SN is
about $5-10\%$ (\citealt{Thornton+98}), thus both the SNIa and
SNII efficiency are assumed to be $\epsilon_{SN}=0.1$. Small
variations (e.g. $\epsilon_{SN}=0.2$) do not affect the results. A
\cite{Salpeter55} initial mass function (IMF), constant in time in
the range $0.1-50 M_{\odot}$, is assumed. Such IMF has been shown
to reproduce the photochemical properties of elliptical galaxies
(e.g., \citealt{Pipino_Matteucci04}; \citealt{Pipino+08};
\citealt{Pipino+10}). A top-heavier IMF would increase the
production of SNe and enhance the SN feedback
(\citealt{Romeo+08}), while a bottom-lighter IMF (as, e.g.
\citealt{Kroupa01}) would not impact SN feedback. Although the
investigation of the effects of the IMF on the model behaviors is
beyond the aims of this paper, the later discussion about the IMF
non--universality (\citealt{Treu+10};
\citealt{Conroy_vanDokkum12b}; \citealt{Cappellari+12};
\citealt{TRN13_SPIDER_IMF}) represents a further important
ingredient of the galaxy evolution to take into account in future
simulations. We will limit to discuss the possible effect of the
IMF variation with galaxy mass in Sects. \ref{sec:discussion} and
\ref{sec:conclusions}, when we will compare our models with real
data.

\subsubsection{Initial conditions}

In Table \ref{tab:SIMs} we show relevant properties of the 4
models we analyze, we have called \Eone, \Etwo, \Ethree\ and
\Efour. In particular, we show both a few important input
parameters - that set the initial conditions and that we discuss
below - and predicted quantities such as the stellar mass and
\rhalf , as well as the metallicity and age gradients. We culled
these four models from those run by \cite{Pipino+08, Pipino+10} to
efficiently scan the space of the setup parameters, i.e. the
central gas density $\rho_{\rm 0,gas}$, the initial profile of gas
distribution, the SF parameter \eSF, and the initial DM content,
$M_{\rm DM}$, that bracket a range of possible initial
configurations in terms of amount of gas initially available,
timescale of accretion/cooling of such gas and timescale of gas
consumption due to star formation. Here, we briefly discuss these
input parameters in detail.
\begin{itemize}
\item $\rho_{0,gas}$.
The values for $\rho_{0,gas}$ are set in order not to limit the
amount of gas in the grid, i.e. smaller than the typical baryon
fraction in high density environment (\citealt{McCarthy+07}).
\Etwo\ has a $\rho_{0,gas}$ of $10$ times smaller than the one in
the other models.
\item {\it Initial profile}. The gas can be initially distributed
as an isothermal sphere (flagged as
\emph{IS}, mimicking a fast accretion of gas before the bulk of
the SF begins) or a uniform distribution within the whole box
(models flagged as \emph{flat}). In the case \Eone\ an {\it IS}
profile is adopted, while for the other models we use the
\emph{flat} profile \footnote{$\rho_{0, gas}$ depends on the
initial profile of gas distribution, being the central density in the
case of {\it IS} profile and constant across the galaxy for the
{\it flat} one.}.
\item \eSF . The SF parameter, \eSF\ (Eq.~(\ref{sfr});
\citealt{Pipino+08}) is taken constant as a function of radius. We
assume values of $1$ and $10$, which guarantee SF rates of 10-500
$M_{\odot}$/yr in massive galaxies, comparable with the
observations of high redshift star forming objects. Models \Eone\
and \Ethree\ have $\eSF = 1$, while for \Etwo\ and \Efour\ is
$\eSF = 10$.
\item $M_{\rm DM}$. The adopted DM masses are $M_{\rm DM} = 2.2 \times 10^{12} \, \Msun$
and $M_{\rm DM} = 5.7 \times 10^{12} \, \Msun$. The DM potential
has been evaluated by assuming a distribution inversely
proportional to the square of the radius at large distances
(\citealt{Silich_Tenorio-Tagle98}). And all these quantities have
been chosen to ensure a final ratio between the mass of baryons in
stars and the mass of the DM halo of around 0.1. The model \Efour\
have the largest initial DM mass. A limitation of the model is
that we assume a fixed DM potential. On the other hand, the
details of the potential are not relevant for the baryonic
processes as long as the DM has a characteristic radius much
larger than that of the baryons (e.g. \citealt{Martinelli+98}).
Moreover, it is still a problem for simulations to reproduce SF
histories that fit the chemical abundance data, if the baryon
accretion follows the DM growth (\citealt{Colavitti+09};
\citealt{Pipino+09}). A more sophisticated
description of the DM will be included in future works.
\end{itemize}
We do not discuss variation in the actual starting value of the
temperature, since it is less important than, e.g., the chosen
$\epsilon_{SF}$ (\citealt{Pipino+10}). Pairs of models differ for
a single parameter, such as by confronting models one-to-one will
allow us to investigate the effect of the individual parameters on
the simulation set-up. In particular, \Eone\ and \Ethree\ have the
same initial conditions but a different gas profile, i.e. {\it IS}
and {\it flat}, respectively. \Etwo\ and \Efour\ have the same
initial gas distribution and \eSF, but a different gas density in
the core. \Ethree\ and \Efour\ have the same $\rho_{\rm 0,gas}$
and gas distribution, but \Efour\ has a larger \eSF.

As extensively discussed in \cite{Pipino+10}, different - but
reasonable - combinations of the input parameters lead to model
properties that reproduce both the overall chemical properties
observed in local elliptical galaxies. Using [$<O$/Fe$>$] as
tracer of [$<\alpha $/Fe$>$] (see Table \ref{tab:SIMs}),
\cite{Pipino+10} have shown that the models adopted hereafter are
satisfactorily consistent with the observations at fixed galaxy
mass (\citealt{Thomas+07}). Indeed, these models also reproduce
the lack of radial gradient in [$<\alpha $/Fe$>$]. At the same
time, these simulations reproduce the mass-metallicity relation
(\citealt{Thomas+07}) and the relation between local escape
velocity and local metallicity (\citealt{Scott+09}). Using colour
gradients we will further constrain the space of physical
parameters involved.

\begin{figure*}
\psfig{file=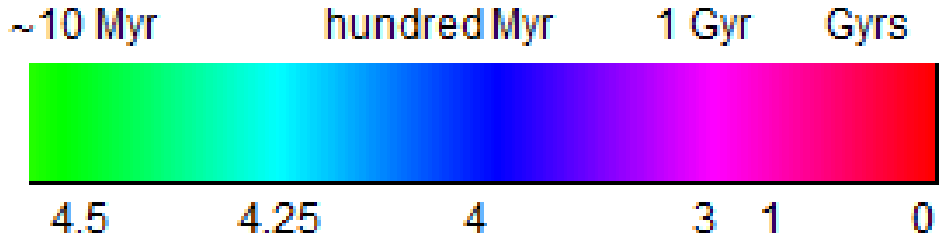, width=0.6\textwidth}\\
\psfig{file=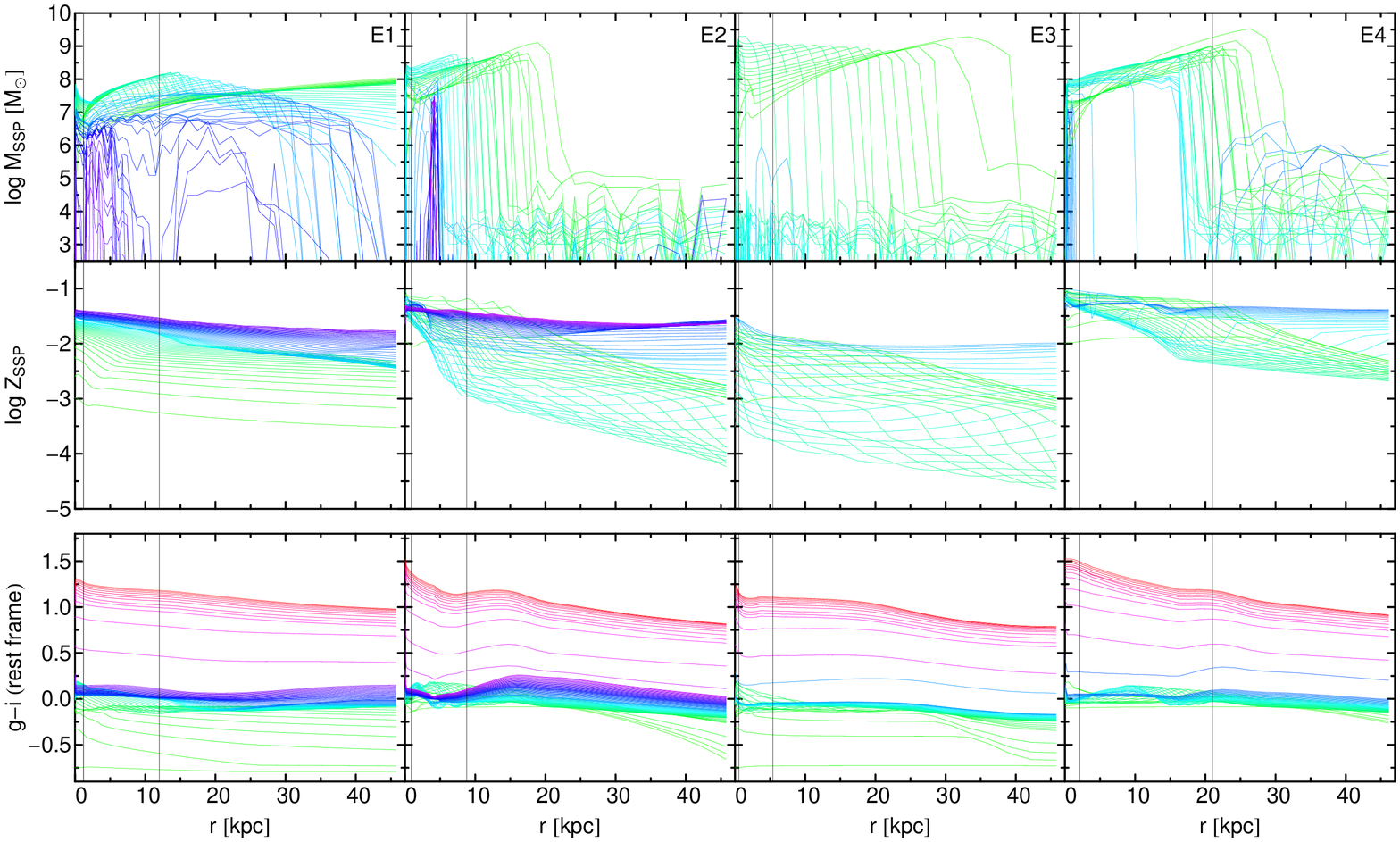, width=0.99\textwidth}\caption{Stellar mass
(top panels) and relative metallicity (middle panels) in SSP
formed at each time step and \gi\ colour (bottom panels), as a
function of the distance from the center of galaxy. From the left
to right \Eone, \Etwo, \Ethree\ and \Efour\ are shown. The
vertical lines set the final $0.1 \rhalf$ and \rhalf. The colour
bar on the top sets the colour code of the curves in terms of the
redshift.}\label{fig:figprofiles}
\end{figure*}

\subsubsection{A brief sketch of the model behaviour}

Before discussing the details of the models it is worth briefly
reminding their general behaviour, whereas we refer the reader to
\cite{Pipino+08} for a comprehensive analysis. At times earlier
than few hundred Myrs the gas is accumulating in the central
regions where the density increases. The temperature drops due to
cooling and, thus, the SF can proceed at very high rates. After a
few hundred Myrs the gas, being heated by SN explosions, is
outflowing at the largest radii, while still being accreted in the
central regions. Afterwards (at times $t_{\rm gw} > 0.2 - 1.2$
Gyr, depending on the model) a galactic wind driven by SNIa+II
involves the entire galaxy. Such a  behaviour is simply explained
by the fact that the  energy required to extract the gas from the
galaxy outskirts is less than the work needed to have an outflow
originating in the galactic center, leading to the correlation
between metallicity and escape velocity gradient, as shown in
\cite{Pipino+10}. This strong wind can be maintained for several
Gyrs , contributing to the ejection of the chemical elements into
the surrounding medium (\citealt{Pipino+02}).

These models exhibit metallicity
gradients in the range $-0.2, -0.4$ dex per decade in radius and
negligible age gradients (cfr. Table \ref{tab:SIMs}). Changes in
the initial conditions within the same broad formation scenario
create the scatter in the predicted gradients at a single mass.

The shallower gradients predicted by this new monolithic models
with respect to earlier dissipational collapse models (e.g.
\citealt{Larson74}; \citealt{Carlberg84}) derives from a
constraint that was not available when the original collapse
models were put forward, namely that it is necessary to produce
stars with high [$<\alpha $/Fe$>$] ratios inhabiting the galactic
core. This turns into a need for short duration of the SF
(accordingly to observations; \citealt{Matteucci94};
\citealt{Thomas+05}). Hence, the metal enrichment process cannot
last more than 1 Gyr, leading to small age gradients even in the
most massive galaxies (e.g. \citealt{Tortora+10CG}).

\begin{figure*}
\psfig{file=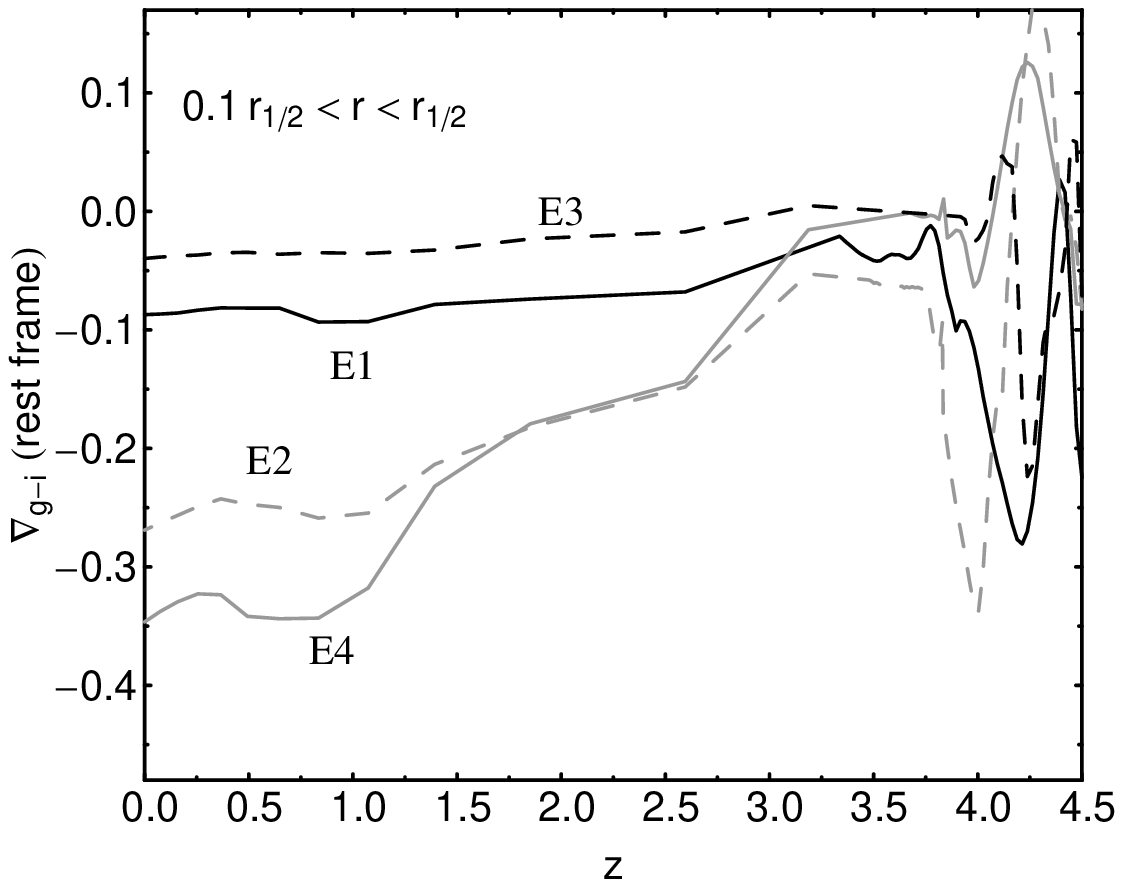, width=0.31\textwidth}
\psfig{file=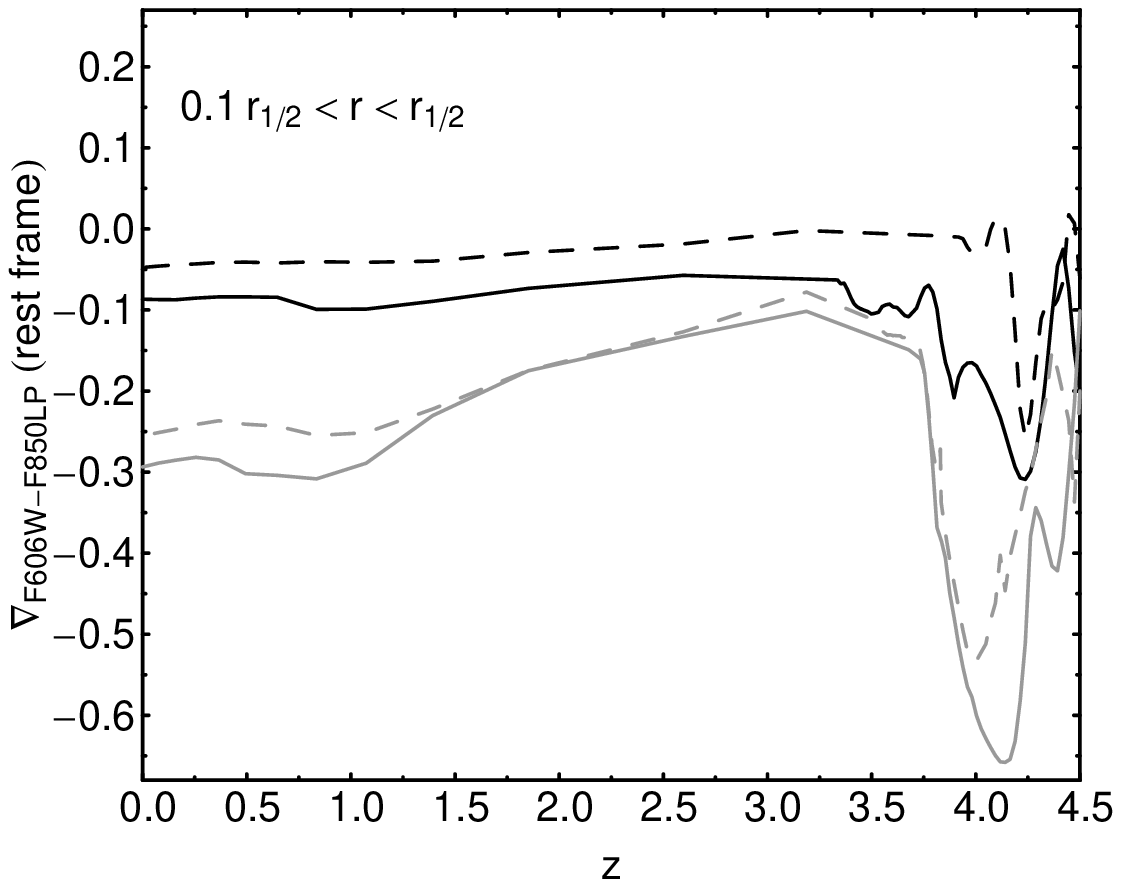, width=0.31\textwidth}
\psfig{file=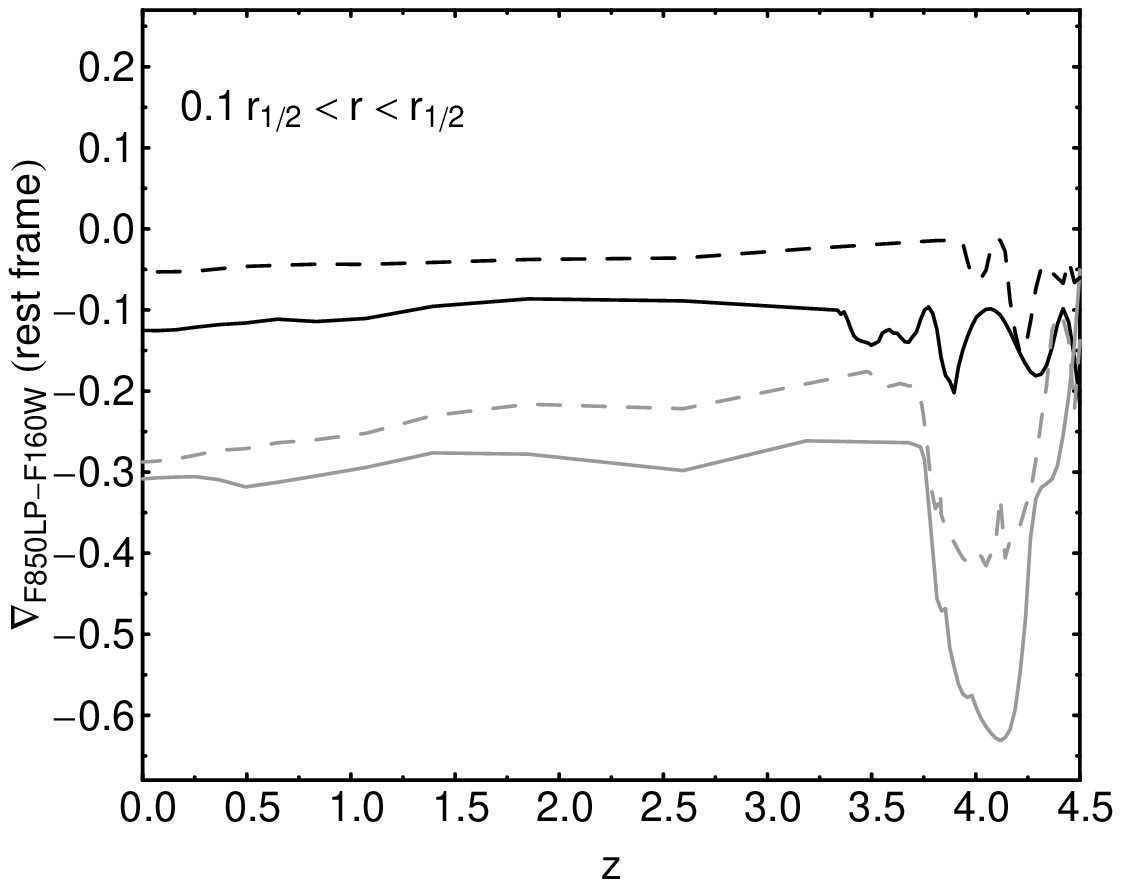, width=0.31\textwidth}\\
\psfig{file=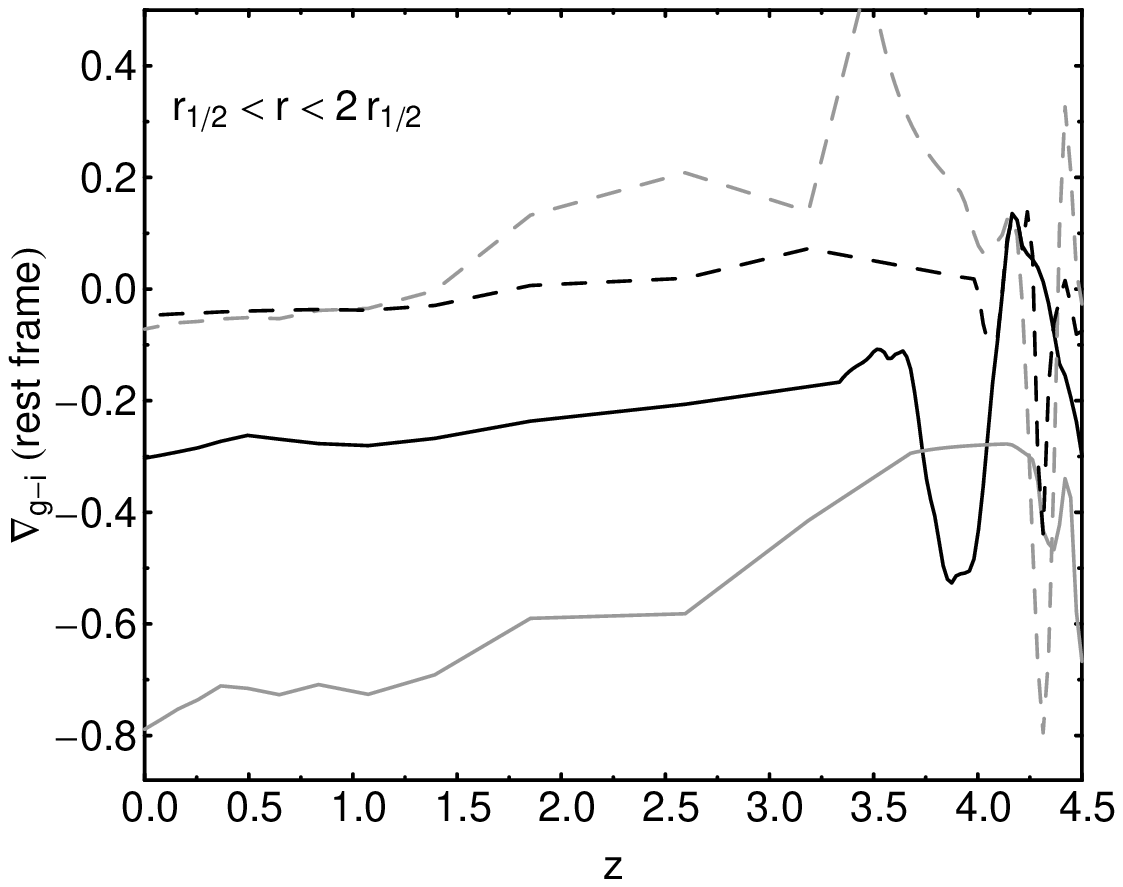, width=0.31\textwidth}
\psfig{file=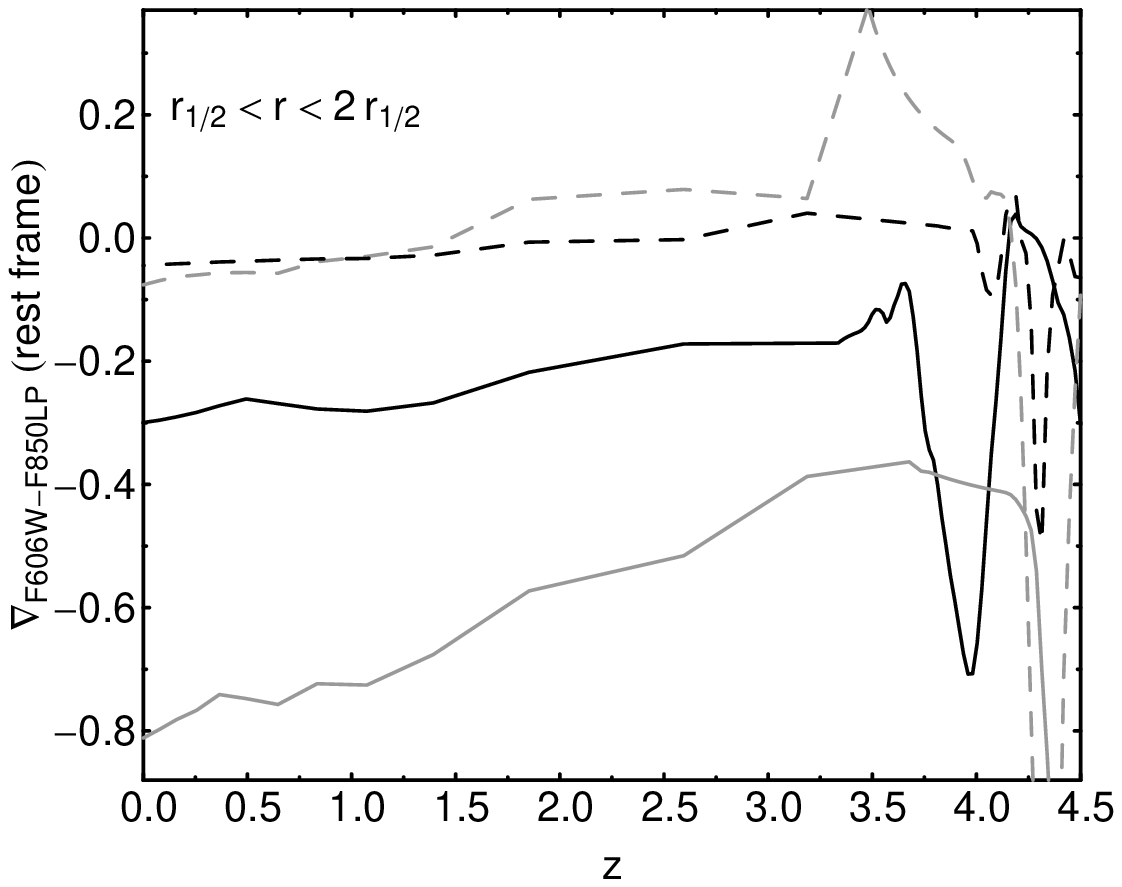, width=0.31\textwidth}
\psfig{file=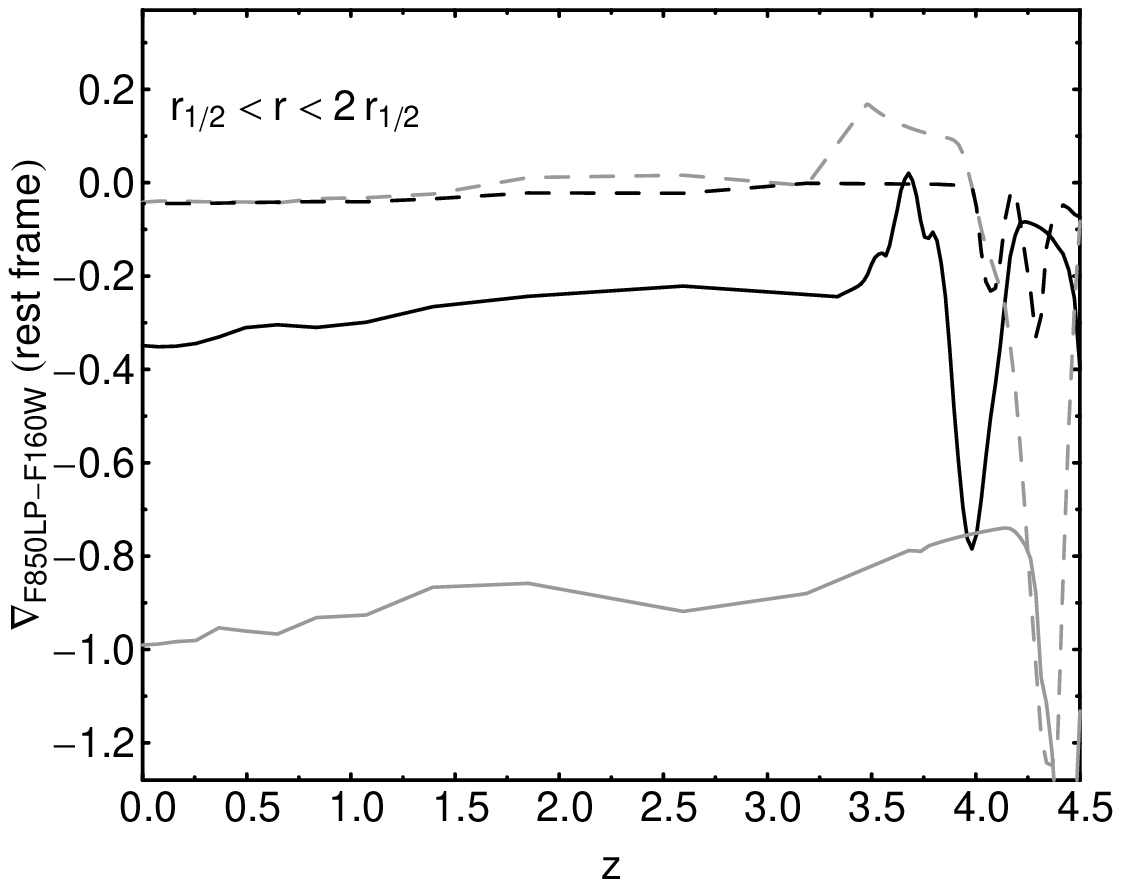, width=0.31\textwidth} \caption{Gradients in
the rest frame of the galaxy as a function of z for \Eone\ (black
solid line), \Etwo\ (gray dashed line), \Ethree\ (black dashed
line) and \Efour\ (gray solid line). From the left to the right
\ggi, \gVIb\ and \gIIRb\ are plotted. In the top (bottom) panels
the gradients are calculated assuming $0.1 \rhalf < r < \rhalf$
($\rhalf < r < 2\rhalf$). }\label{fig:gradients_rest_frame}
\end{figure*}

\subsection{Stellar population analysis}

To derive quantities that can be compared with observations we use
a set of ``single burst'' stellar population synthesis (SPS)
spectra from \citet[hereafter BC03]{BC03} as the building blocks
of our composite SPs. To match the assumptions in the simulations
a \cite{Salpeter55} IMF is adopted, metallicity, $Z$, ranging
within $0.0001 - 0.05$ and ages, t, in the interval $0.001-15 \,
\rm Gyr$ are used. To improve the sensitivity of the results to
the small differences in both $Z$ and ages, we have adopted a
``mesh refinement'' procedure which interpolates the synthetic
models.

For each galaxy, and at each radius, we discretize the star
formation history. Specifically, we link a BC03 SED to the $i-th$
SSP formed at a time $t_{i}$, with a given $Z_{\rm SSP}$ and a
mass $M_{\rm SSP}$, assuming that the burst that created it is
instantaneous. We assume that at times $t>t_{i}$ it passively
evolves till present day. Thus, at each time and radius, we have a
collection of SPS which we weigh according to the mass $M_{\rm
SSP}$ assembled in every previous timestep, and sum to derive the
composite galaxy spectrum at that fixed time and spatial position.
Suitably convolving the spectral response of filters with the SED,
we derive the surface brightness and magnitudes. In the present
paper we discuss the results for several SDSS and HST pass bands,
which will be compared with observations from the literature. We
apply this procedure to the SSPs formed in spherical annuli with
radii from 0 to $\sim 45$ kpc, to reconstruct the colour profiles
and average colours. To investigate spurious systematics arising
from the assumption of a specific SPS prescription we have also
adopted an updated version of BC03 models (BC03-updated,
hereafter; \citealt{Eminian+08}) and \citet[M05
hereafter]{Maraston05} models, which include an improved treatment
of the thermally-pulsing asymptotic giant branch phase (TP-AGB).

We discuss the colour profile $(X-Y)(r)$ where $X$ and $Y$ are the
two bands considered. The colour gradient is defined as the
angular coefficient of the relation $X-Y$ vs $\log r$,
$\displaystyle \nabla_{X-Y} = \frac{\delta (X-Y)}{\delta \log r}$,
measured in $\rm mag/dex$ (omitted in the following unless needed
for clarity). To take the analysis as simple as possible we avoid
any projection of galaxy light distribution, calculating the
gradients using the 3D distribution and the spherical half-mass
radius \rhalf. We have verified that for the scope of this paper
the impact of this assumption is small, leaving unaffected our
conclusions. In the worst cases, the projection will flatten the
steepest gradients of, at most $0.1-0.2$ mag/dex. The fit of each
color profile is performed in different radial ranges; in
particular, to homogeneously compare with observations we often
use the range $0.1 \rhalf \leq r \leq \rhalf$.

The spherical assumption is a possible limitation for our models,
since calculating the colour gradients of more realistic triaxial
or oblate systems along different lines of sights would produce
different gradients. The inclusion of more complex geometries is a
natural step forward in this kind of analyses. For the present
paper, we have limited our models to be as simple as possible,
demonstrating that they can satisfactorily reproduce key
low-redshift observations (\citealt{Pipino+08, Pipino+10}). Under
the assumption that the galaxy formation is fairly homogeneous, as
we expect to be in the case of an isolated galactic system, the
presence of asymmetries in mass distribution is expected to be
marginal. In this case it is instructive to study if any observed
variation in the gradient can be traced back to variations in
physical quantities, before taking the further step of relaxing
the assumption of galaxies evolving homogenously and in isolation
and introducing the effect of the merging on the gradient
distribution.  We note that, even in this last case,
\cite{Hopkins+09_DELGN_II} have demonstrated that gradients along
different lines of sights of remnants of disk galaxies mergers
present a scatter around the median gradient which is less that
0.1, namely smaller than the observed scatter at a fixed galaxy
mass.

On the observational side, while it is true that age/metallicity
profiles taken at two different position angles may differ (e.g.
Fig. 7 in \citealt{CDB93}), and we are aware that inhomogeneities
and structures are clear also in 2-dimensional metallicity maps of
ellipticals (e.g. \citealt{Kuntschner+06}), these are plausibly
second-order effects, related to the particular assembly history
of a given galaxy and contributing very little to the scatter in
the data.

\section{Colour gradients and evolution with
redshift}\label{sec:CG_evolution}

In this section, we discuss the main predictions of our simulations,
concentrating on the evolution of colour profiles and colour
gradients in the rest-frame of the galaxy, which provide us with
information about the physics behind the simulations and the link
with observables.

The profiles for the mass formed, metallicity and \gi\ color for the
four models are shown in Fig. \ref{fig:figprofiles}, whereas the
gradients \ggi, \gVIb\ and \gIIRb\ as a function of z are plotted
in Fig. \ref{fig:gradients_rest_frame}. For all the models and at
each epoch the galaxies are predicted to be bluer in the outer
regions, with some exceptions at very high redshift, where
positive gradients can be found. In the following we list the main
characteristics of the models, discussing the main features in the
profiles and concentrating on the colour gradients.

The early evolution of model \Eone\ is characterized by a rapid
accumulation of stellar mass in the outskirts, which is then
halted by the development of winds. At the same time, gas is still
being accreted in the central regions. The faster speed of the
metal enrichment process (due to star formation) relative to the
dilution (given by the inflow of pristine), gas makes the gas
metallicity steadily increase with time. On the other hand, for
the  models \Etwo, \Ethree\ and \Efour, the winds start to expel
outer gas soon, stopping SF in the outer regions. Afterwards,
winds are continuously activated in more and more central regions,
and the centers accrete pristine/metal poor gas from outer
regions, which, during the initial phases, dilutes the metals
produced in situ. At later stages, new SPs are formed by gas
processed by stars and start to be metal-richer.

\begin{itemize}
\item \Eone. The galaxy is initially forming a larger
fraction of stars in the outer regions (till $\sim 45 \, \rm
kpc$).  Since SN feedback occurs externally and then internally,
at later stages of the evolution the trend with radius of the mass
formed present a peak and decline at large radii. All the new SSPs
formed are metal-richer. The metallicity profile declines with the
radius at all the redshifts and presents steeper gradients in the
central regions with respect to the outer ones, with the radius
separating the two regimes increasing with the time. The inflow
rate is slower than SF and outflow rates, therefore the new gas
accreted, in quickly enriched in metals. The interplay between the
stellar mass build-up and the evolution in metallicity produces
declining \gi\ profiles. In particular, the bending in the colour
profile is driven by the bending in the Z profiles. If we consider
the gradients (black solid line in Fig.
\ref{fig:gradients_rest_frame}) in the range $0.1 \rhalf < r <
\rhalf$, in the early stages ($z \sim 4, 4.5$) the gradient is
steep ($\ggi \sim -0.25$, and shallower in the other colours), but
during the next evolution the gradients are almost unchanged and
$\sim -0.1$, independently of the colour adopted. The gradients
are steeper if we consider a more external region $\rhalf < r < 2
\rhalf$, where, similarly to the inner region we find very steep
gradients at early stages and a flattening at $z \lsim 4$. After
this initial phase, the galaxy evolves for passive evolution and
the gradients are affected by a slight steepening setting on
values of $\sim -0.35$ at the present day.
\item \Etwo. In each time interval, the mass formed in SSP
is smaller at smaller radii, increasing at $10-15$ kpc and
dropping in the outermost regions. As the time passes the radius
where the $M_{\rm SSP}$ becomes null gets smaller. The effect is
the same of \Eone\ and is caused by SN feedback. The Z profile
created by the initial burst of star formation, is initially flat
in the central regions and decreasing outside. Due to the large
amount of inflowing gas the SSPs formed at slightly later times
incorporate less metals than those of the previous generation. At
the same time, the central regions start to show a steeper
gradient than the outer ones. The metallicity reaches a minimum
and then increases again due to the high SF parameter adopted in
this model. The winds start to expel gas soon in the outskirts,
quenching the SF in these regions. Afterwards, winds are
continuously activated in more and more central regions.
Contemporarily, the centers accrete pristine/metal poor gas from
outer regions, which, during the initial phases, dilutes the
metals produced in situ. At later stages, new populations of stars
are formed by gas processed by stars and start to be metal-richer.
The combination of these trends gives initially a gradient which
is steeper, and then get flatter, with the colour staying
constant, the subsequent passive evolution reinforces a bump in
the colour profile at $\sim 15 \, \rm kpc$ and a very steep
gradient in the central regions ($\lsim 7-8 \, \rm kpc$). In the
range $0.1 \rhalf < r < \rhalf$, the gradients (gray dashed line
in Fig. \ref{fig:gradients_rest_frame}) present very steep values
and get shallower at $z \sim 3.5$. Passive evolution makes \ggi\
steeper in time ($\sim -0.25$), while the steepening is weaker in
the other colours. In the range $\rhalf < r < 2 \rhalf$, this
model predicts strong negative gradients at $z\sim 4.3$, which
become null and then positive till $z \sim 1.5$, this feature is
weaker in the \gIIRb. They become almost null at the present
epoch.
\item \Ethree. The star formation history is similar to the one described for the
previous models, but the SF is more extended in radius. The
metallicity evolution is also quite similar to \Etwo, with
metal-poorer SSP as the time increases and an inversion of this
trend with metal-richer SSPs at later stages. The colour profile
is flat at high redshift, with steeper profiles in the outer
regions. At later stages the profile is steep in the very central
regions ($\lsim 1 \, \rm kpc$), stays flatter in the central
regions ($\lsim 15 \, \rm kpc$), declining in the very outer ones.
The evolution of the gradient (black dashed line in Fig.
\ref{fig:gradients_rest_frame}) in the range $0.1 \rhalf < r <
\rhalf$ is quite similar to \Eone, but the gradients are
shallower. The same similarity is not found in the region $\rhalf
< r < 2 \rhalf$ since the model presents a negative spike at $z
\sim 4.3$ (similar but less pronounced than for \Etwo), and a
gradient which independently of the colour remains very shallow
($\sim 0$) till the present day.
\item \Efour. The mass formed is
less extended in radius, staying in between \Etwo\ and \Ethree.
Although the average larger metallicity at all the epochs, the
evolution is the same of \Etwo\ and \Ethree. The net result in the
colour is similar to \Etwo. At very high redshift the colour
profile has a particular wavy shape and at later stages evolve
similarly to \Etwo, with the formation of a bump at $r \sim 20 \,
\rm kpc$. The \ggi\ (gray solid line in Fig.
\ref{fig:gradients_rest_frame}) in the region $0.1 \rhalf < r <
\rhalf$ show some positive and negative values at higher redshift,
while the other two colours only show very steep negative
gradients ($\gVIb \sim \gIIRb \sim -0.6$), but after this
turbolent initial phase, at $z\sim 3$ the gradients becomes
flatter and while \ggi\ and \gVIb\ steepen with time, \gIIRb\
stays constant, similarly to the other models. In the range
$\rhalf < r < 2 \rhalf$, after \gVIb\ and \gIIRb\ being very steep
and negative at $z \sim 4.3$, the gradients become shallower at $z
\sim 4$ and again steeper with the time ($\ggi \sim \gVIb \sim
-0.8$ and $\gIIRb \sim -1$ at $z=0$).
\end{itemize}

\begin{figure}
\centering \psfig{file=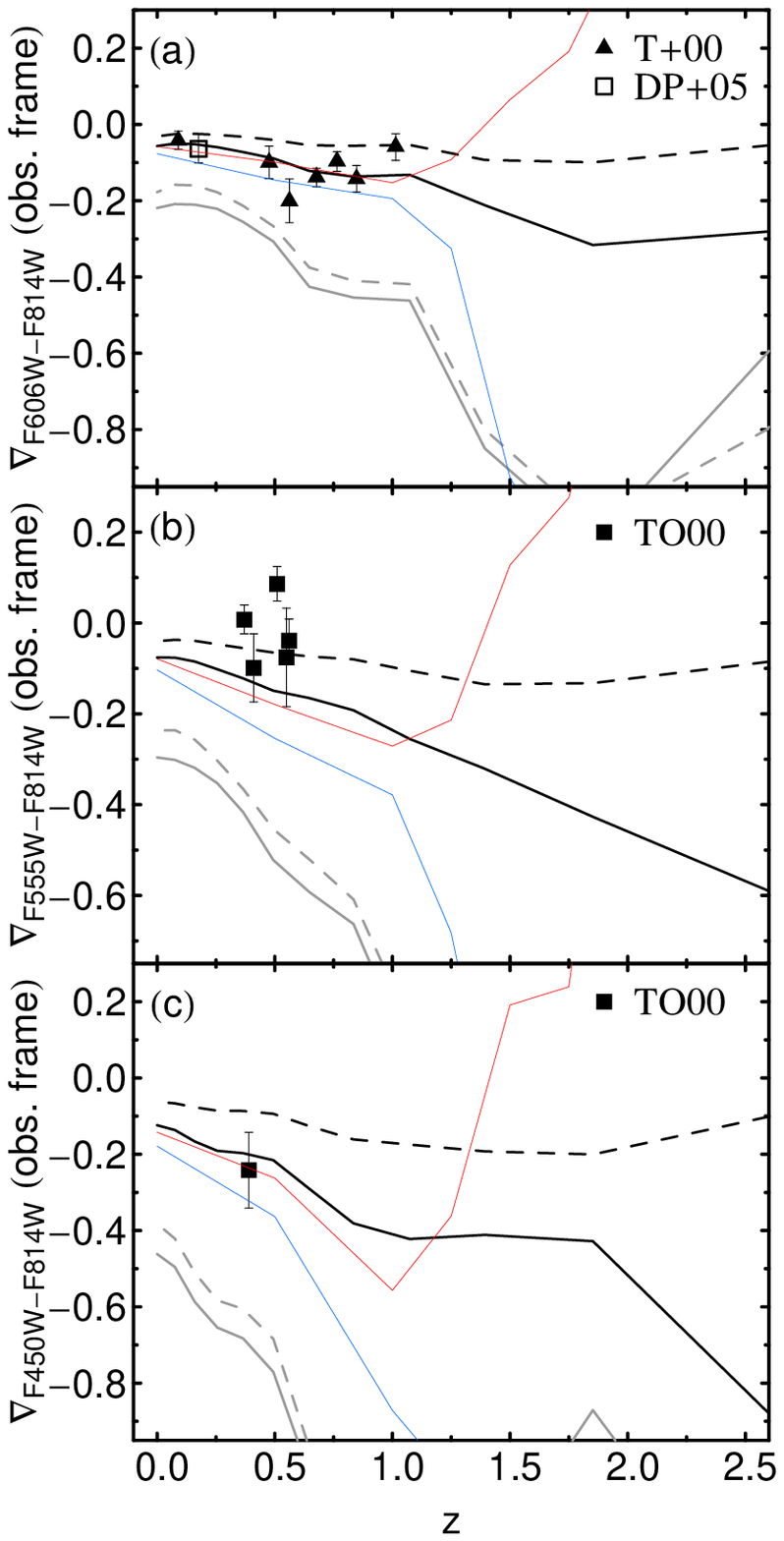, width=0.365\textwidth}
\caption{Evolution of observer frame colour gradients and
comparison with observations. From top to bottom we plot \gVIa,
\gVIc\ and \gVId. The meaning of the curve-styles is as in Fig.
\ref{fig:gradients_rest_frame}. The literature data used for the
comparison are from \citet[T+00]{Tamura+00},
\citet[DP+05]{DePropris+05} and \citet[TO00]{Tamura_Ohta00}. We
also plot two toy-models obtained using BC03 synthetic models: 1)
the blue line is for the model with a metallicity gradient of $\gZ
\sim -0.4$ constant with redshift and a $z=0$ age gradient of
$\gage \sim 0.04$, 2) the red line is for $\gZ \sim -0.4$ and a
$z=0$ $\gage \sim - 0.04$.}\label{fig:gradients_obs_frame}
\end{figure}

It is now interesting to link the differences found in the
gradient normalizations and in the trends with redshift to the
initial conditions of the models. In particular, we discuss in the
following the relative role of these differences.
\begin{itemize}
\item {\it Initial profile.} \Eone\ and \Ethree\ have a different
initial gas profile, {\it IS} and {\it flat}, respectively. The
{\it flat} profile in \Ethree\ gives shallower gradients and the
difference is larger in the external region. Bluer total colours
are produced.
\item $\rho_{0, gas}$. The role of the central gas density can
be analyzed by comparing the predictions for the models \Etwo\ and
\Efour . It has a small role in the central regions ($0.1 \rhalf <
r < \rhalf$), but in the outer region ($\rhalf < r < 2 \rhalf$) it
has a strong impact on the gradients, since the smaller central
density in \Etwo\ produces the positive colour gradients at high
redshift and the very flat ones at $z \lsim 2$. \Etwo\ has also a
bluer total colour (bluer than observations). Note that for models
\Etwo\ and \Efour\ (and similarly for \Ethree), the gas profile is
flat, i.e., the gas density is $\rho = \rho_{0, gas}$ everywhere,
thus it is not surprising that $\rho_{0, gas}$ affects the outer
regions, where it can be immediately consumed to formed stars,
more than central ones, where new gas is continuously accreted.
\item {\it \eSF .} \Ethree\ and \Efour\ have a different \eSF\
parameter. In both the regions, the gradients are steeper in
\Efour, thus the higher the SF parameter, the steeper the
gradient. From \cite{Pipino+10}: basically, an increase in the SF
parameter enhances the differences between the inner core and the
outskirts set by the other initial conditions.
\end{itemize}

\section{Comparison with observations}\label{sec_observations}

We now compare the predictions from our simulations with
observations. We have collected several samples ETGs, with stellar
masses comparable to ours, at intermediate redshift ($z<1$) and
higher redshift ($1 < z < 2.5$).
\begin{itemize}
\item In Sect. \ref{sec:lowz} we start comparing observer frame gradients with low/intermediate redshift galaxies
(\citealt{Tamura+00}; \citealt{Tamura_Ohta00};
\citealt{DePropris+05}).
\item In Sect. \ref{sec:highz} we compare observer- and rest-frame
colour gradients with high redshift observations
(\citealt{Guo+11}; \citealt{Gargiulo+11, Gargiulo+12}).
\end{itemize}

\begin{figure}
\centering \psfig{file=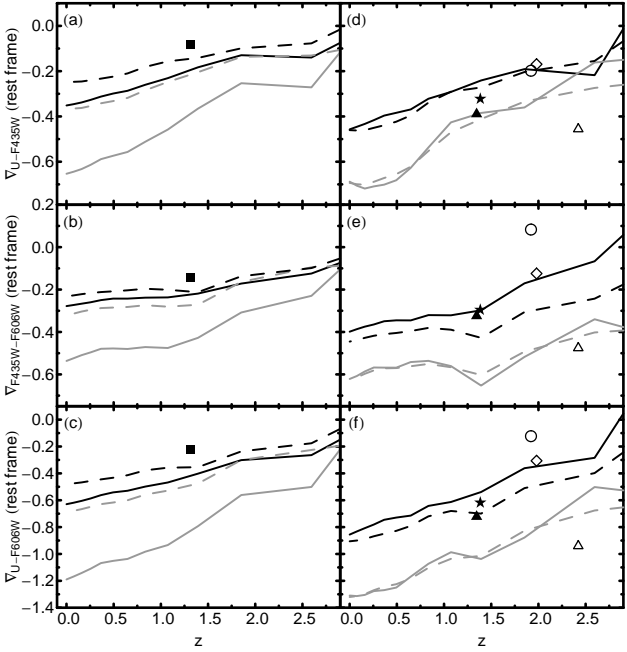, width=0.49\textwidth}
\caption{$\nabla_{\rm U-F435W}$, $\nabla_{\rm F435W-F606W}$ and
$\nabla_{\rm U-F606W}$ in the rest-frame of the galaxies as a
function of redshift. The symbols are as in Figs.
\ref{fig:gradients_rest_frame} and \ref{fig:gradients_obs_frame}.
In panels (a), (b) and (c) the gradient is calculated at $r
> 0.5 \, \rhalf$, while in panels (d), (e) and (f) at $r > 1.5 \, \rhalf$.
The data points are for galaxies in \citet{Guo+11}: GOODS IDs
24626 (full box), 19389 (full triangle), 22704 (full star), 23555
(open circle), 24279 (open square) and 23495 (open triangle). For
24626, the authors provide the colour gradients down to $\sim 0.5$
times the effective radius, while for the other galaxies the
gradients are calculated down to $\sim 1.5$ times the effective
radius. The uncertainties on the gradients are $\sim 0.2$ mag/dex,
the error bars are omitted for
clarity.}\label{fig:gradients_vs_Guo}
\end{figure}

\subsection{Low/intermediate redshift
observations}\label{sec:lowz}

Simulations are compared with data at $z<1$ in Fig.
\ref{fig:gradients_obs_frame}. In panel (a) we compare the model
\gVIa\ with archival Hubble Deep Field North (HDFN) data of 7
bright ETGs up to $z=1$ from \cite{Tamura+00} and the average
gradient (with 1$\sigma$ scatter) of the 22 ETGs in Abell 2218
analyzed in \cite{DePropris+05}. We find a fairly good agreement
with our \Eone\ model, \Ethree\ is consistent with the highest
redshift galaxy.

In panel (b) the model \gVIc\ are compared with gradients for ETGs
in 5 clusters at redshifts $z=0.37-0.56$ ($\rm A370$, $\rm CL\,
0939+47$, $\rm CL\, 0412-65$, $\rm CL\, 0016+16$ and $\rm CL\,
0054-27$) with measured photometry in F555W and F814W band passes.
The average gradients with $1 \sigma$ scatter in each cluster are
plotted. The agreement is worse with \Eone\ except for $\rm CL
0939 + 47$ and data seem to better agree with \Ethree .

In panel (c) the model \gVId\ are compared with the mean gradients
for ETGs in $\rm CL 0024 + 16$ analyzed in \cite{Tamura_Ohta00}.
Here the agreement with \Eone\ is very good.

In general, $z<1$ data seem to favor models where the gradient is
shallow, close to $z\sim 0$ values, without strong signs of
evolution\footnote{Similarly to the rest-frame evolution in Fig.
\ref{fig:gradients_rest_frame}. However, although the filters
sample different parts of the spectra at different redshifts, at
lower redshift the impact of this effect is small.}. This is not
entirely surprising, if one sets the formation of ETGs at $z>2$
with passively evolving galaxies already at $z\sim2$. However,
half of our models (i.e. \Etwo\ and \Efour), despite the high
formation redshift and the passive evolution, still predict a
strong evolution of the color gradients (see the strong rest-frame
evolution in Fig. \ref{fig:gradients_rest_frame}).

To ease the comparison between such observations and model
predictions in Fig. \ref{fig:gradients_obs_frame} we have also
plotted two toy-model galaxies adopting the BC03 SPS, with
negative metallicity gradient of $\gZ \sim -0.4$, and small
positive (negative) age gradients of $\gage \sim 0.04$ ($\gage
\sim -0.04$) at $z=0$. In these models, the \gZ\ is constant as a
function of redshift (since the Z is fixed in the BC03 models),
while \gage\ is changing, getting steeper and steeper at higher
redshift. Following the redshift evolution we notice that the two
toy-models reproduce fairly well observations at $z < 1$
(consistently with our monolithic predictions).

Thus, we have shown that the SP gradients predicted by our models
(for instance \Eone\ and \Ethree, i.e. with final $\gZ \sim -0.2$
and $\gage \sim 0$) reproduce local and $z<1$ observations.
Consistently with the toy-models plotted in Fig.
\ref{fig:gradients_obs_frame} our conclusions are in agreement
with \cite{Tamura+00} and \cite{Tamura_Ohta00}, which have
demonstrated that metallicity gradients alone do reproduce the
observed colour gradients (with null age gradients), whereas age
gradient alone do not.

\subsection{High redshift observations}\label{sec:highz}

A larger scatter in the color gradients is, instead, expected at
$z>1$, when we approach the epoch of formation of the galaxies and
even a small scatter in their assembly redshift, or the earlier
phases of passive evolution, may enhance the radial differences in
the SP properties.

\begin{figure}
\centering \psfig{file=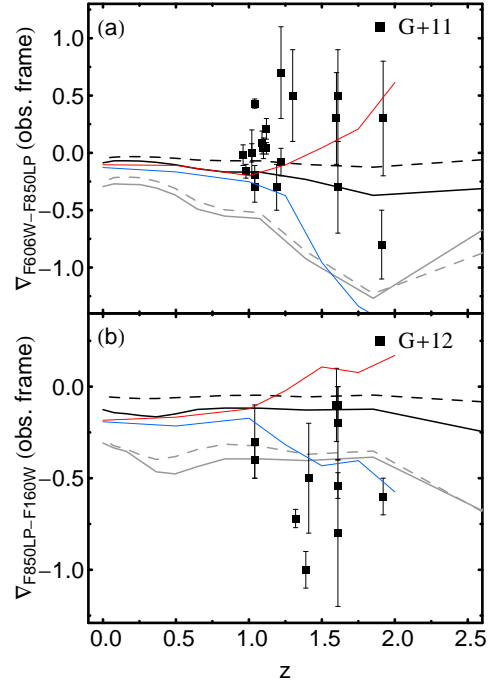, width=0.38\textwidth}
\caption{The same as Fig. \ref{fig:gradients_obs_frame}. From top
to bottom we plot \gVIb\ and \gIIRb. The literature data used for
the comparison are from \citet[G+11]{Gargiulo+11} and
\citet[G+12]{Gargiulo+12}. The style of the curves and toy-models
are as in Fig.
\ref{fig:gradients_obs_frame}.}\label{fig:gradients_obs_frame_bis}
\end{figure}

In Fig. \ref{fig:gradients_vs_Guo} we compare our predictions with
rest-frame colours from \cite{Guo+11}. \cite{Guo+11} have analyzed
the properties of six massive galaxies (with masses in the range
$\sim 1.5-13 \times 10^{10} \, \Msun$, assuming a Salpeter IMF)
from the GOODS south field in VIMOS-VLT U-band, and HST F435W- and
F606W-bands. They have first converted their data to rest-frame
quantities, measuring gradients from $\sim 1.5$ to $\sim 8$ times
the H-band effective radius, and in one case (ID 24626) they were
able to follow the gradients down to $\sim 0.5$ times the
effective radius. To homogenously compare with these data, in Fig.
\ref{fig:gradients_vs_Guo} we calculate the gradients from $0.5$
(left panels) and $1.5$ (right panels) times the \rhalf\ of the
model galaxy. Although such observations have been converted to
rest-frame quantities, introducing further uncertainty to the
comparison, we find an excellent agreement with our set of models,
with only a couple of outliers.

Moreover, in Fig. \ref{fig:gradients_obs_frame_bis} the model
predictions for \gVIb\ and \gIIRb\ are compared with the results
for ETGs at $1 < z < 2$ selected from the GOODS-South field and
analyzed in \cite{Gargiulo+11} and \cite{Gargiulo+12},
respectively. These galaxies are selected on the basis of a visual
inspection of F850LP images and of the S\'ersic index, $n_{\rm
F850LP}$ ($n_{\rm F850LP} > 2$), and span a range of masses $\sim
(3-50)\times 10^{10}\, \rm \Msun$ (Salpeter IMF). From the panel
(a), within the data and simulation uncertainties \Eone\ and
\Ethree\ are quite consistent with a large fraction of the ETGs.
Unfortunately, our simulations are not able to reproduce the
positive \gVIb\ found in some of the galaxies, since negative \gZ\
and negligible \gage\ are produced at all the epochs (see Table
\ref{tab:SIMs}). On the contrary, from the panel (b), few
datapoints agree with \Eone\ and \Ethree\ within uncertainties,
some other galaxies are in better agreement with \Etwo\ and
\Efour\ and few other ones have quite steep gradients the models
do not reproduce.

In Fig. \ref{fig:gradients_obs_frame_bis} we have also plotted the
two toy-model galaxies first shown in Fig.
\ref{fig:gradients_obs_frame}. At $z \gsim 1$, the strongest (with
respect to $z<1$) positive and negative \gage\ produce a strong
evolution of colour gradients. A discrepancy emerges from these
results, since strong positive \gage\ at large redshifts fit the
positive $F666W-F850LP$ gradients found in \citet[panel
(a)]{Gargiulo+11}, while a negative \gage\ would reproduce the
steep negative $F850LP-F160W$ gradients in \citet[panel
(b)]{Gargiulo+12}.

\subsection{Systematics and missing ingredients}\label{sec:syst}

\begin{figure*}
\centering \psfig{file=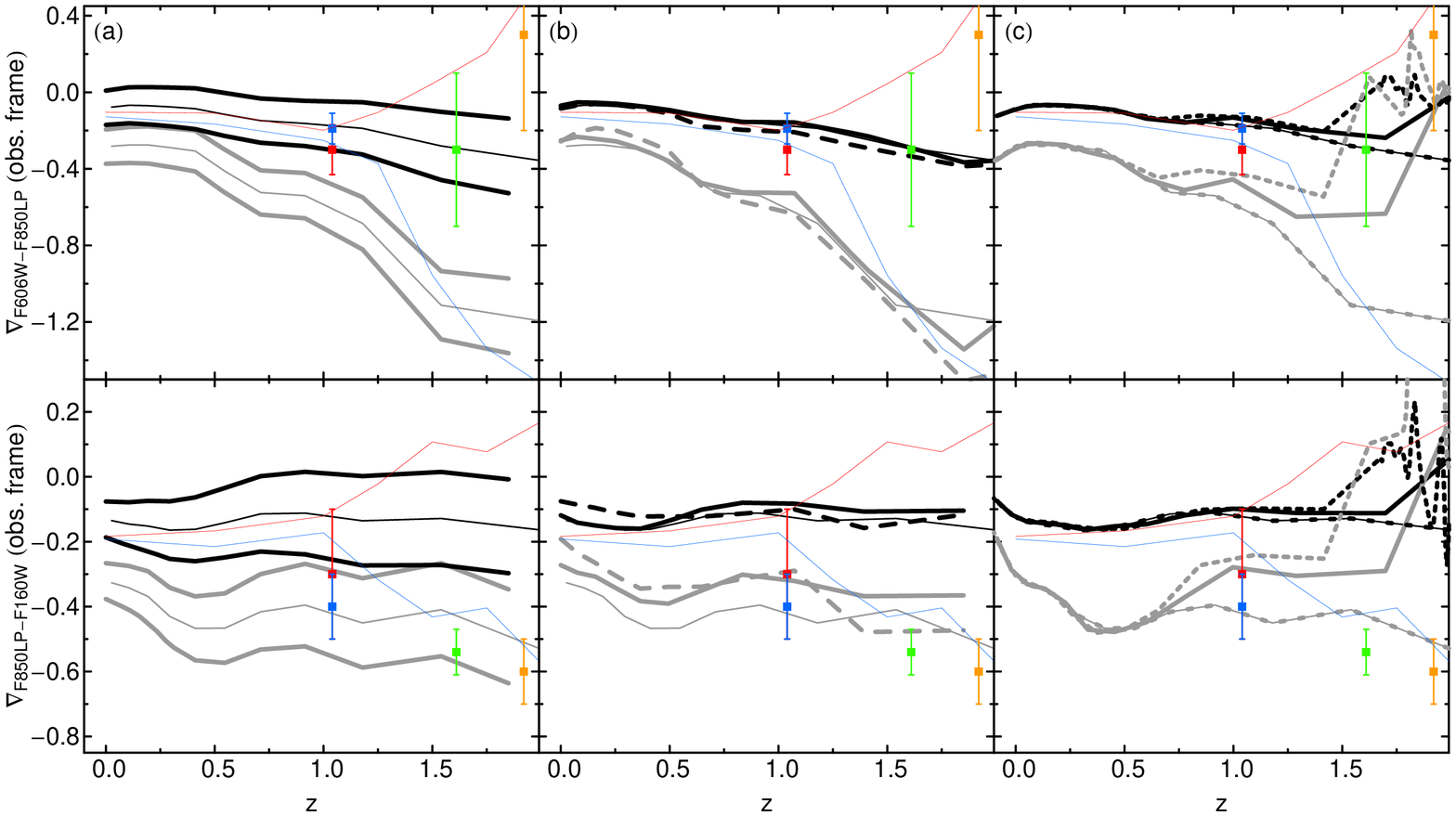, width=0.9\textwidth}
\caption{Systematics in observer frame colour gradients. \gVIb\
(top panels) and \gIIRb\ (bottom panels) for \Eone\ (black lines)
and \Efour\ (gray lines) are plotted as a function of z. The
reference model is plotted as thin line. From the left to the
right we show: (a) the role of positive and negative gradients in
the extinction (which bracket the case with no dust extinction);
(b) the role of SPS prescription, BC03 is compared with
BC03-updated (thick line) and M05 (dashed thick line); (c) the
impact of a change of the formation redshift: $z_{\rm f} = 2.5$
(thick line), $2$ (short dashed thick line) and $6$ (long dashed
thick line); see text for further details. The red and blue lines
are the same toy-models plotted in Figs.
\ref{fig:gradients_obs_frame} and
\ref{fig:gradients_obs_frame_bis}. The colored boxes and bars are
for the four galaxies shared by \citet{Gargiulo+11} and
\citet{Gargiulo+12}: IDs 23 (red), 11888 (blue), 2111 (green) and
472 (orange).}\label{fig:gradients_obs_frame_2}
\end{figure*}

In order to investigate further the discrepancies discussed above
we have:
\begin{itemize}
\item[a)] analyzed the impact of some ingredients, (as
dust extinction, SPS prescription and formation redshift) on our
results, and
\item[b)] performed a more detailed comparison with
high-redshift observations, picking those galaxies in common between
the two papers (\citealt{Gargiulo+11, Gargiulo+12}), having both
$F666W-F850LP$ and $F850LP-F160W$ measured (i.e. the galaxies with
IDs 23, 11888, 2111 and 47; see the papers above for further
details).
\end{itemize}

\subsubsection{Systematics}

We start by analyzing the role of dust extinction, SPS
prescription and formation redshift on our results in Fig.
\ref{fig:gradients_obs_frame_2}. For sake of clarity we only limit
the analysis to the models \Eone\ and \Efour \footnote{From Fig.
\ref{fig:gradients_obs_frame} we see that the predicted gradients
calculated in the range $0.1 \rhalf < R < \rhalf$ for \Eone\ and
\Ethree\ are similar, and the same holds for \Etwo\ and \Efour .
For this reason, for the scope of our comparison we can limit the
analysis to \Eone\ and \Efour, and extend the general
considerations to the other models.}.

We first show in panels (a) the role of dust extinction. We adopt
the extinction law from \cite{CCM89} with $R_{V}=3.1$, calculating
the corrections to the observer-frame colours and the relative
colour gradients. Due to our ignorance about dust gradients in
high redshift galaxies, we will rely on the simplest hypothesis,
we will assume a dust extinction gradient of $\nabla E(B-V) = \pm
0.2$ constant with redshift\footnote{The only relevant quantity
here is the variation of the extinction in terms of the radius
(i.e. the gradient $\nabla E(B-V)$) and not the absolute values of
$E(B-V)$ at each radius. Moreover, we exclude steeper dust
gradients, which would be unrealistic for central galactic regions
($\lsim \rhalf$ as it is in Fig. \ref{fig:gradients_obs_frame_2}).
If a $\nabla E(B-V)$ changing with redshift were adopted, then the
outcoming gradient evolution would lie within the ranges
plotted.}. The impact on the gradients is of $\pm 0.1$, a steeper
(shallower) negative gradient corresponds to the model with a
larger (smaller) fraction of dust in the center.

In panels (b) we show that the use of either the BC03-updated
prescription (\citealt{Eminian+08}) or the M05 models leaves
almost unchanged the optical colour gradient, while can provide
shallower near-IR gradients.

The effect of the change in the formation redshift is shown in
panels (c), where we notice that smaller formation redshifts (e.g.
$z_{\rm f} = 2$ and $2.5$) give shallower colour gradients at $z
\gsim 0.5$ and positive at $z \gsim 1.5$.

\subsubsection{Understanding the discrepancies with high-z ETGs}

In Fig. \ref{fig:gradients_obs_frame_2} we also plot the four
galaxies shared by \cite{Gargiulo+11} and \cite{Gargiulo+12}. The
galaxy with ID 23 agrees with \Eone, within statistical
uncertainties. This agreement improves if we have more dust in the
center than in the peripheries, making the gradients slightly
steeper. If we consider more dust in the peripheries the galaxy
also matches the results from \Efour. However, this scenario seems
not plausible (\citealt{Gargiulo+12}). The agreement with \Efour\
can be improved if we assume a smaller $z_{\rm f}$ for our model.
Consistently with our models, a synthetic toy-model with a
suitable negative \gZ\ and a mild (positive or negative) age
gradients can also reproduce the observations for this galaxy (see
also Fig. \ref{fig:gradients_obs_frame}). For the other galaxies
(IDs 11888, 2111, 472) no agreement can be found with our models
and synthetic toy-models explored. The situation gets worse for
galaxies 2111 and 472. These systems are characterized by null or
positive \gVIb\ and steep negative \gIIRb, which cannot be
accounted by our models, and nor by any combination of \gage, \gZ,
gradients in the dust extinction, a change in $z_{\rm f}$. As a
confirmation of our results, by means of a SPS analysis
\cite{Gargiulo+12} reach similar conclusions, suggesting the
simultaneous variation of several parameters (e.g., age, Z, SF
duration, IMF slope, dust extinction, etc.), which will require
further analysis.

\section{Physical scenario}\label{sec:discussion}

\begin{figure}
\centering
\includegraphics[trim= -1mm 0mm 5mm 0mm, width=0.43\textwidth,clip]{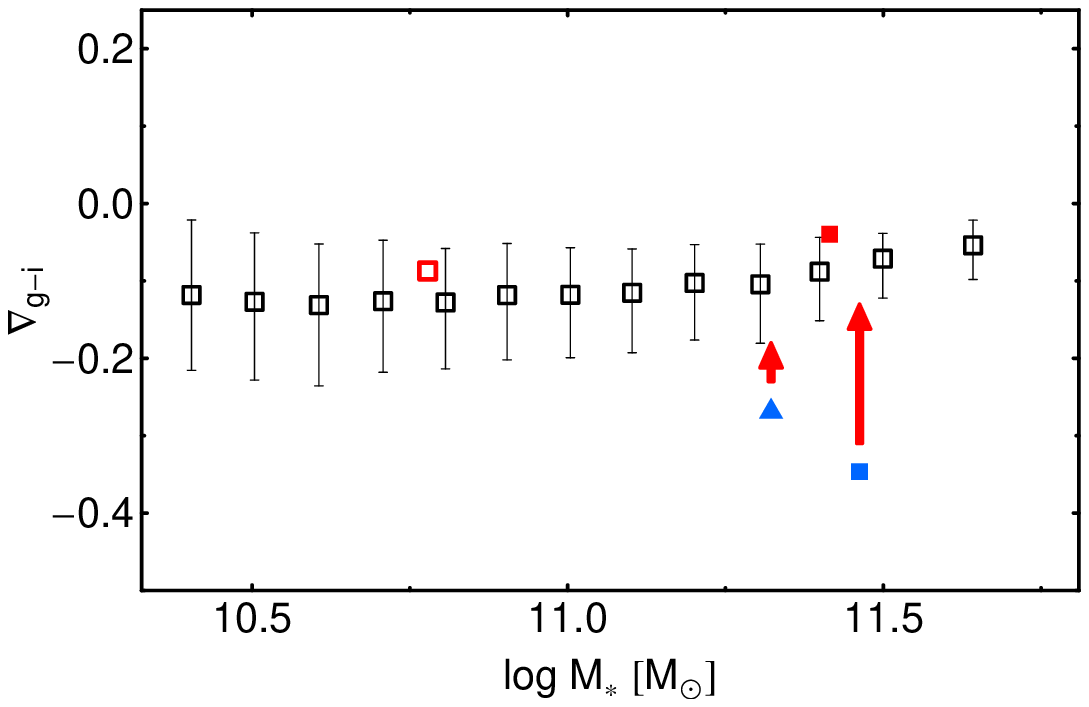}\\
\includegraphics[width=0.45\textwidth,clip]{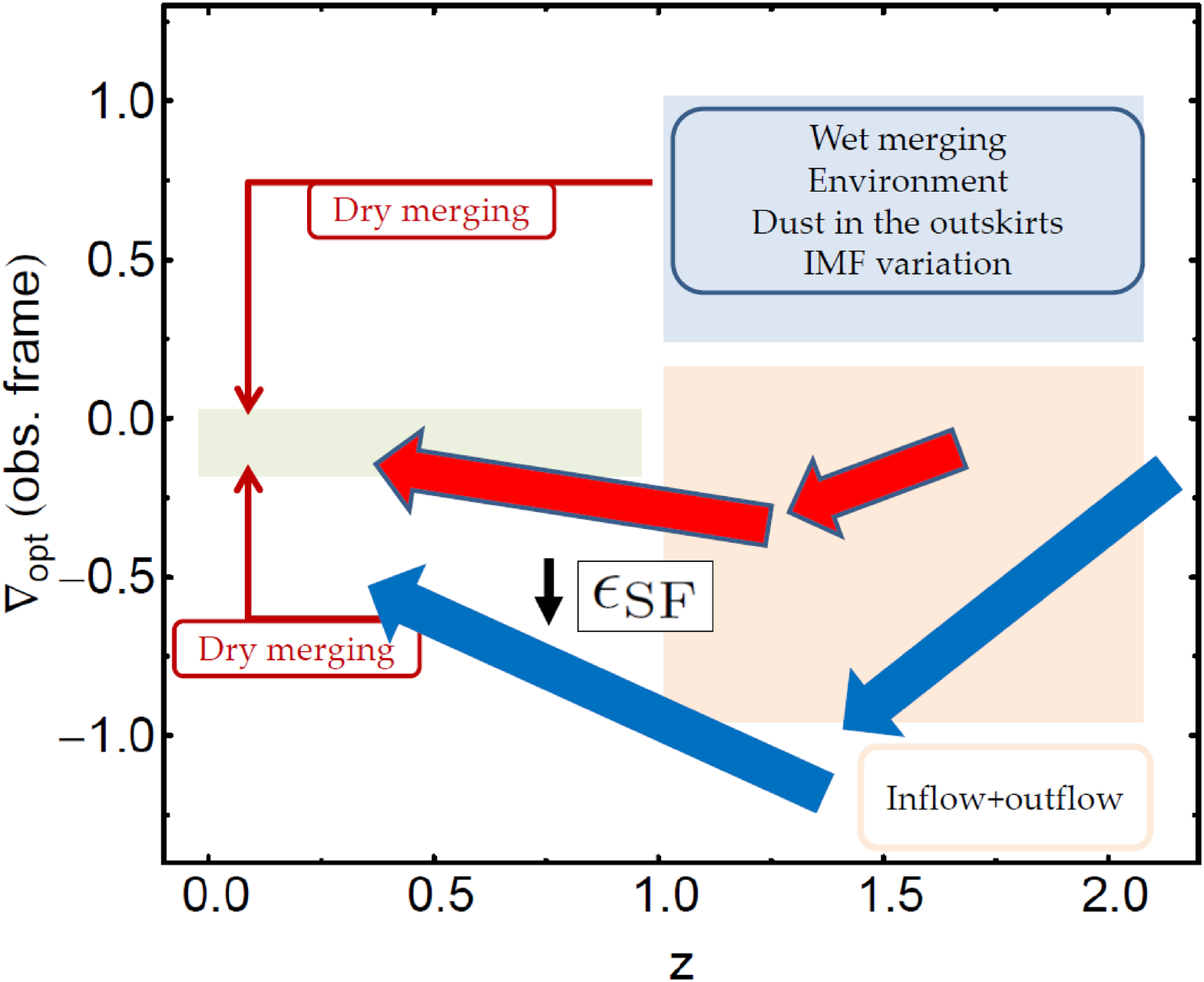}
\caption{Top panel. \ggi\ as a function of stellar mass. We plot
the $z=0$ gradients for \Eone\ (red open box), \Etwo\ (blue filled
triangle), \Ethree\ (red filled box) and \Efour\ (blue filled
box). The black open boxes and error bars are for medians and
25-75th quantiles of results for local SDSS ETGs in
\citet{Tortora+10CG}. The red vertical arrows show a qualitative
impact of dry mergings on the model gradients. Bottom panel.
Optical colour vs redshift in the observer frame. The shaded
rectangles set the regions on the plane where observations fall.
The red and blue arrows give a qualitative view of the trends of
our models (red arrows for \Eone\ and \Etwo; blue arrows for
\Etwo\ and \Efour) as shown, for instance, more quantitatively in
Figs. \ref{fig:gradients_obs_frame} and
\ref{fig:gradients_obs_frame_bis}. The black small arrow show how
an increase of SF parameter affects gradients. Information about
the physical processes involved are also shown. See the text for a
full discussion of this picture.}\label{fig:cartoon}
\end{figure}

We finally gain insights on the physical processes or any missing
ingredients in our analysis. The general physical picture is given
in the bottom panel of Fig. \ref{fig:cartoon}. We have shown that,
although there are some disagreements with some high redshift
observations (e.g. \citealt{Gargiulo+11, Gargiulo+12}), our
monolithic models can reproduce a wide collection of
low/intermediate redshift data samples and a fraction of data at
high redshift (e.g. \citealt{Tamura+00}; \citealt{Tamura+00};
\citealt{DePropris+05}; \citealt{Guo+11}). The interplay between
the evolution of the metallicity profiles and the radial behaviour
of the SF history, both combined via the SPS prescription, drives
the evolution of the optical and near-IR colours. Stars in all
models are formed at very early stages ($z_{\rm f} \gsim 4$) and
in a short period, in agreement with previous literature
(\citealt{Matteucci94}; \citealt{Thomas+05}), while at later
stages the galaxies evolve passively. The optical colours
typically show stronger redshift evolution during this passive
phase when compared with the near-IR one. In the early stages the
models also present some strong features and very strong negative
colour gradients.

Thus, accordingly to other sets of observables (as metallicity and
$[\alpha/Fe]$), a wide range of observations can be reproduced by
the interplay of gas infall and outflow from evolved stars and
supernovae. In particular, while the models with a smaller \eSF\
can consistently reproduce the data at $z \lsim 1$
(\citealt{Tamura+00}; \citealt{Tamura+00}; \citealt{DePropris+05})
and local observations (see top panel in Fig. \ref{fig:cartoon};
\citealt{Tortora+10CG}), those with larger \eSF\ produce colour
gradients which are steeper than local observations. To reconcile
the latter models with results at $z \lsim 1$ further physical
ingredients are necessary. The occurrence of at least one dry
major merger at $z \lsim 1$ (\citealt{DiMatteo+09}) could wash out
the gradients from the latter models (\Etwo\ and \Efour) producing
shallower gradients which are consistent with local estimates (see
top panel in Fig. \ref{fig:cartoon}; \citealt{Kobayashi04}) and
reproducing quite well all the other local correlations, but the
occurrence of such events seem relatively uncommon
(\citealt{DePropris+10}).

Our models have been built to reproduce the local negative
metallicity gradients at low redshift and do not predict positive
observer-frame \gVIb\ at high redshift (jointly with the negative
\gIIRb). Such positive gradients, although not common in the local
massive ETGs, are found at higher redshift (e.g.
\citealt{Menanteau+01}; \citealt{Ferreras+09_GOODS};
\citealt{Gargiulo+11}). They can be possibly reconciled with the
simulated negative gradients if we assume a) they have a huge
amount of dust in the external regions ($\nabla E(B-V)
> 0.2$; Fig. \ref{fig:gradients_obs_frame_2}), and/or b) interactions with environment have stripped off the outer
gas, and/or c) a wet merging has generated a starburst in the core
(\citealt{Hopkins+09_DELGN_II}), and/or d) a change of IMF with
radius inducing positive colour gradients (\citealt{Gargiulo+12}).
Subsequent dry mergings could wash out such gradients contributing
to match the local observations. If on the one hand too strong
gradients in the dust seem unrealistic, our models are built for
isolated galaxies, and do not take into account exogen phenomena
as merging/close encounters/etc. or any variation in the IMF,
which could make redder the outskirts and the observed gradients
positive. However, it seems not realistic to reproduce some
gradients observed (\citealt{Gargiulo+11, Gargiulo+12}) with any
combination of age, metallicity and dust extinction (see Fig.
\ref{fig:gradients_obs_frame_2}). Further physical ``unexplored''
galactic ingredients accounting for such features need to be
investigated. Recently, many works have contrasted the IMF
universality, suggesting a systematic IMF variation with mass
(\citealt{Conroy_vanDokkum12b};
\citealt{Cappellari+12_ATLAS3D_XX}; \citealt{SPIDER-VI};
\citealt{Ferreras+12}; \citealt{TRN13_SPIDER_IMF}) and with
redshift (\citealt{vanDokkum08}). It is likely that IMF can change
as a function of redshift and within each galaxy, inducing non
trivial variations in colour gradients (\citealt{Gargiulo+12}).

\section{Conclusions}\label{sec:conclusions}

We have adopted monolithic hydrodynamical models of four massive
ETGs (which we have called \Eone, \Etwo, \Ethree\ and \Efour) from
\cite{Pipino+08}, which are combined with SPS models from
\cite{BC03} to convert the simulated metallicities and SF history
into observed quantities as colour gradients. Therefore, we
provide a suite of model predictions which can be directly
compared with low/high redshift measurements (without any
intermediate step, as the SPS fitting to derive age/metallicity
gradients in measured colour gradients). So far the new monolithic
models proposed and analyzed in \cite{Pipino+08} and
\cite{Pipino+10} -- characterized by two main contrasting
phenomena, infall of gas from the outer regions, and outflows due
to the stellar and SN winds -- reproduce some global correlations
in local galaxies, as the Z- and $[\alpha/Fe]$-mass relations and
give shallower \gZ\ with respect to earlier models (e.g.
\citealt{Larson74}; \citealt{Carlberg84}). In particular,
\cite{Pipino+10} have demonstrated that the models reproduce
fairly well a) the average Z gradients ($\gZ \sim -0.3$) and b)
the observed scatter in local ETGs. This improvement of the models
has been possible thanks to the requirement, matched by the
models, that the stars produced have high average $[\alpha/Fe]$ in
the cores and thus short SF (as found by observations;
\citealt{Matteucci94}; \citealt{Thomas+05}; \citealt{Tortora+09})
and smaller age gradients (\citealt{Tortora+10CG}). The aim of the
present paper has consisted to present the redshift evolution of
the color gradient in the simulated galaxies and to further
constrain the model parameters by means of new and independent
dataset. From the comparison with data, we infer the following
main conclusions:

\begin{itemize}
\item The models \Eone\ and \Ethree\ show a quite good agreement with
data sets at  $z \simeq 0$ (local
SDSS galaxies from \cite{Tortora+10CG} - the top panel of Fig.
\ref{fig:cartoon}), $z \lsim 1$ (Fig. \ref{fig:gradients_obs_frame_2};
\citealt{Tamura+00}; \citealt{Tamura_Ohta00};
\citealt{DePropris+05}) as well as with the rest-frame colour
gradients derived in \cite{Guo+11} for a sample of six
high-redshift ($z > 1$) galaxies.

As already pointed out in \cite{Pipino+10} (for the intrinsic metallicity
gradients) the higher is the SF parameter,
the steeper the color gradient. Since the observations agree with those
models with shallower gradients (\Eone\ and \Ethree), the data-model
comparison seems to favor small values for \eSF.

\item Despite the previous
encouraging results, some disagreement is found when we further add the
high redshift galaxies by \citealt{Gargiulo+11, Gargiulo+12}. In
particular, we find that the optical data tend to
favor models producing the shallower colour and metallicity
gradients (i.e. \Eone\ and \Ethree), while most of the
near-infrared gradients are consistent with the models with
steeper gradients (i.e. \Etwo\ and \Efour).

\item Assuming that this latter case is true, models with higher \eSF\ would require the
intervention of some process which flatten the gradients, as, at
least, 1 dry merger at $z \lsim 1$ to match local observations.

\item Models allow us to investigate the role of other physical processes,
parameterized through different initial conditions, as the central
gas density, the gas density distribution. Their effect on the
gradient is smaller than that of \eSF, therefore, we present
predictions that can be constrained when the statistics on high
redshift gradients improve. In particular, when the initial gas
distribution is flat, then the colour gradients are shallower and
the difference is stronger in the outer regions. A smaller gas
density, as in the model \Etwo\ will give slightly shallower
colour gradients in the central regions ($0.1 \rhalf < r <
\rhalf$), while positive (at $z \gsim 1.5$) or null (at $z \lsim
1.5$) in the outer region ($\rhalf < r < 2 \rhalf$).

\item To further investigate the inconsistency among the data in
\cite{Gargiulo+11} and \cite{Gargiulo+12} we have picked out the
four galaxies with both the F606W-F850LP and F850LP-F160W colours
measured, and we find that our models can reproduce consistently
only one of the selected galaxies, having some troubles to fit the
others. \cite{Gargiulo+12} have pointed out that the variation of
a single SP parameter is not able to reproduce the observed colour
gradients and global SED. This would suggest that a simultaneous
variation of several parameters has to be invoked to reproduce
such a data. Unfortunately, we have verified that, not only our
models do not work, but, using SPS toy-models, combinations of age
and metallicity gradients fail to reproduce such observations.
Some further parameter or more complex phenomena would be taken
into account to reproduce such observations.
\end{itemize}

We have finally discussed our results within a more complex
physical scenario (see Sect. \ref{sec:discussion}). As we have
amply discussed, although our models reproduce several
observational data, seem to fail in some galaxies at $1 < z < 2$.
In particular, our models, by construction, do not reproduce the
steep positive optical colour gradients (\gVIb) in
\cite{Gargiulo+11}, which can be possibly originated by a) wet
mergings, b) interaction with environment, c) strong dust
extinction gradients or d) IMF variation across the galaxies.
Moreover, the unsuccessful matching of combined data from
\cite{Gargiulo+11} and \cite{Gargiulo+12} seem to be a difficult
task to achieve adopting combinations of gradients in age,
metallicity and dust extinction, suggesting that more work is
needed to reproduce observations, including further and unexplored
ingredients in the simulations. Larger observational samples of
gradients in high-redshift ETGs (out to one effective radius and
beyond) can confirm the results discussed in this paper,
validating the models or suggesting the intervention of further
physical processes, as (dry or wet) galaxy mergings, environmental
phenomena, AGN feedback or questioning the assumptions about
stellar ingredients, as the IMF universality suggesting possible
variations in terms of the galactocentric radius
(\citealt{TRN13_SPIDER_IMF}).


\section*{Acknowledgments}

We thank the anonymous referee for the useful suggestions which
helped to improve the paper. CT was supported by the Swiss
National Science Foundation and the Forschungskredit at the
University of Zurich. AD was supported by PRIN-INAF 2011 "Multiple
populations in Globular Clusters: their role in the Galaxy
assembly". FM acknowledges financial support from PRIN
MIUR2010-2011, project N. 2010LY5N2T.


\bibliographystyle{mn2e}   


\end{document}